\documentclass[]{interact}

\usepackage{epstopdf}
\usepackage{subfigure}

\usepackage[numbers,sort&compress,merge]{natbib}
\usepackage{endnotes,refcount}
\bibpunct[, ]{(}{)}{,}{n}{,}{,}
\renewcommand\bibfont{\fontsize{10}{12}\selectfont}
\renewcommand\citenumfont[1]{\textit{#1}}
\renewcommand\bibnumfmt[1]{(#1)}

\theoremstyle{plain}

\theoremstyle{definition}

\theoremstyle{remark}

\begin{document}

\articletype{ARTICLE TEMPLATE}

\title{Studying fundamental physics using quantum enabled technologies with trapped molecular ions}

\author{
\name{D.~M. Segal\textsuperscript{a}$^{,\ast}$\thanks{$^\ast$Deceased.}, V. Lorent\textsuperscript{b}\thanks{CONTACT V. Lorent. Email: vincent.lorent@univ-paris13.fr}, R. Dubessy\textsuperscript{b} and B. Darqui\'{e}\textsuperscript{b}\thanks{CONTACT B. Darqui\'{e}. Email: benoit.darquie@univ-paris13.fr}}
\affil{\textsuperscript{a}Quantum Optics and Laser Science, Blackett Laboratory, Imperial College London, Prince Consort Road,
London SW7 2AZ, United Kingdom; \textsuperscript{b}Laboratoire de Physique des Lasers, Universit\'{e} Paris 13, Sorbonne Paris Cit\'{e}, CNRS, 99 Avenue Jean-Baptiste Cl\'{e}ment, 93430 Villetaneuse, France}
}

\maketitle

\begin{abstract}
The text below was written during two visits that Daniel Segal made at Universit\'{e} Paris 13. Danny stayed at Laboratoire de Physique des Lasers the summers of 2008 and 2009 to participate in the exploration of a novel lead in the field of ultra-high resolution spectroscopy. Our idea was to probe trapped molecular ions using Quantum Logic Spectroscopy (QLS) in order to advance our understanding of a variety of fundamental processes in nature. At that time, QLS, a ground-breaking spectroscopic technique, had only been demonstrated with atomic ions. Our ultimate goals were new approaches to the observation of parity violation in chiral molecules and tests of time variations of the fundamental constants. This text is the original research proposal written eight years ago. We have added a series of notes to revisit it in the light of what has been since realized in the field.
\end{abstract}

\begin{keywords}
Cold molecular ions; tests of fundamental physics; quantum logic spectroscopy; ultra-high resolution molecular spectroscopy; parity violation; variations in the fundamental constants
\end{keywords}

\section{Foreword}

The text below was written during two visits that Daniel Segal made as an invited professor at Universit\'{e} Paris 13. Danny stayed at Laboratoire de Physique des Lasers (LPL) the summers of 2008 and 2009 to participate in the exploration of a novel lead in the field of ultra-high resolution spectroscopy, at that time jointly considered by two teams from LPL: the \emph{M\'{e}trologie, Mol\'{e}cules et Tests Fondamentaux} (\emph{MMTF}) group who address fundamental issues in physics using molecular spectroscopy and the \emph{Bose-Einstein Condensation} (\emph{BEC}) team who investigate quantum degenerate atomic gases in confined geometries. Although the objectives pursued and methods developed by molecular and atomic physicists happen to be quite different, molecules have the potential to improve precise spectroscopic tests of fundamental physics and outperform measurements on atoms. Nevertheless, compared to molecules, the motion and internal state of cold atoms are easier to manipulate and control. In this context, the demonstration of Quantum Logic Spectroscopy on atomic ions by the group of David Wineland in 2005 opened new perspectives. Our idea was to extend this technique to probe trapped molecular ions and carry out precise spectroscopic measurements in order to address open questions in physics. However, the price we had to pay was to start an entirely new challenging experimental project with ions. Danny was the person who could convince us to do so. Under his guidance, the potential of using atomic and molecular ions for precise spectroscopic measurements was explored. This led to a range of valuable observations, from the most fundamental to the most practical. On the fundamental side, for instance, quantum chemistry predicts that chiral ionic molecules may well show a higher sensitivity to parity violation effects compared to neutral species (Trond Saue, Universit\'{e} Paul Sabatier, Toulouse, private communication). On the practical side, we found out that part of the required know-how was existing at LPL and that collaborations with the French ion trapping community were not only feasible, but also welcome.

The text that follows is an original research proposal written eight years ago, of which Danny has been a (if not the) main contributor. It has, in the end, never been submitted to any funding agency, and thus far, we did not take the plunge and did not switch from neutral to ionic species yet. However, we find it interesting to revisit the text in the light of what the community has realized since then. We have thus added notes to the original text accordingly, which we hope will persuade the reader of both the significance of the research proposed here, and the bright and clear-sighted vision that Danny had of his field.

The document is organized as follows. The next section is an introduction that motivates this work in a retrospective way, in light of the state-of-the-art of the field back in 2009. Section 3 and 4 are as originally written by us in 2009. Section 3 summarizes the background needed and gives the state-of-the-art in 2009 regarding ion trapping, cooling and state manipulation, and regarding Quantum Logic Spectroscopy. Section 4 details the envisioned work plan. Added notes update the reader on the state-of-the-art and the evolution of the field since 2009. The overall narrative enriched by the notes therefore happens to be quite unusual for a paper in a conventional journal, but we believe it properly conveys Danny's voice.

\section{Introduction}

Studying the fundamental forces of nature is a core goal of modern physics and has spawned a huge variety of methods and approaches. A general misconception is that in order to perform fundamental tests, enormous collective efforts such as those under way at CERN are the only possible solution. However, the tenacity of atomic, molecular and optical physicists has meant that many vital pieces of the puzzle have been put in place through meticulous research involving nothing more than the interaction of matter with light. Key predictions of quantum electrodynamics (Lamb shift~\cite{Kolachevsky2009}, spontaneous emission~\cite{Jhe1987}) and the Standard Model of particle physics (Parity Non-Conservation in the Electro-Weak interaction, for a review, see~\cite{GUENA2005}) have been elucidated using these methods. Limits have been placed on cosmological models that allow for variations in the fundamental constants. In 2009, a number of groups worldwide were using the methods of quantum optics to search for physics beyond the Standard Model (\emph{e.g.} Supersymmetry) by attempting to measure, for instance, the electric dipole moment (EDM) of the electron~\cite{Regan2002,Hudson2002}\endnote{Since 2009, the research into the electron EDM has seen remarkable advances. Molecules have made a sizeable impact here, as the best limits on the size of the electron EDM are now set using cold YbF~\cite{Hudson2011} and ThO~\cite{Baron2014} molecules, outperforming the limit set by measurements on atoms~\cite{Regan2002}. The present proposal argues that, compared to atoms and neutral species, molecular ions have the potential to improve on precision tests of fundamental physics. Thus, in this context, the recent demonstration, in the group of Eric Cornell, of the first measurement of the electron EDM using trapped molecular HfH$^+$ ions~\cite{Cairncross2017} is of particular pertinence. The resulting electron EDM upper bound is only 1.4 times larger than the current record using neutral species~\cite{Baron2014}, and offers the potential to improve on this limit in the near future.\label{edm}}. The \emph{MMTF}  group at Villetaneuse has played a key role in the general area of optical tests of fundamental physics. In 2008, measurements at LPL, in conjunction with the SYRTE Laboratory (Syst\`{e}mes de R\'{e}f\'{e}rence Temps-Espace, Paris, the French national metrology institute) set the tightest direct and model-free constraints thus far on possible current epoch variations of the electron-to-proton mass ratio~\cite{Shelkovnikov2008}\endnote{Since then, the search for varying fundamental constants, or for putting constraints on their variation, has become a very active part of experimental science. Recently, laboratory constraints on a drift rate of both the fine structure constant, $\alpha$, and the electron-to-proton mass ratio, $\mu$, in the present epoch have been determined by measuring two optical transitions in $^{171}$Yb$^+$ ions~\cite{Godun2014,Huntemann2014}. Assuming a linear drift rate, intertwined constraints of $\dot{\mu}/\mu<10^{-16}$~yr$^{-1}$ and $\dot{\alpha}/\alpha<10^{-17}$~yr$^{-1}$ were derived. The LPL 2008 measurement that resulted in $\dot{\mu}/\mu=(-3.8\pm5.6)\times10^{-14}$~yr$^{-1}$~\cite{Shelkovnikov2008} remains the only current epoch direct $\mu$ constraint derived from a molecular study. As such, although less constraining, it is less model dependent. Complementary astrophysical measurements have also led to stringent constraints on the rate of change of $\mu$ on a cosmological timescale. The most stringent current constraints are derived from methanol absorption lines at redshift $z=0.89$~\cite{Jansen2014,Ubachs2016} and translate into $\dot{\mu}/\mu=(1.4\pm1.4)\times10^{-17}$~yr$^{-1}$, if a linear rate of change is assumed.\label{mudot}}.

Tests of fundamental physics based upon matter-light interactions may also have a bearing on some surprising scientific disciplines. In biology, one of the most fascinating and fundamental open questions concerns the fact that bio-molecules are very often chiral (\emph{i.e.} they have a ‘handedness’) and that furthermore, systems seem to have evolved towards preferentially adopting one chirality over the other. Parity violation has been suggested as a possible root cause of this asymmetry in nature~\cite{TranterA.J.MacDermottR.E.OverillP.J.Speers1992}. Indeed parity violation effects in molecules have been predicted but have, thus far, eluded detection due to their small size. The tightest limits on parity violation effects in chiral molecules have been set by the experiments of the \emph{MMTF} group at the LPL~\cite{Daussy1999,Ziskind2002}, and, in 2009, new measurements were to be performed with neutral molecules seeded in continuous supersonic beams, allowing Doppler-free spectroscopy with interrogation times of the order of a few milliseconds\endnote{The supersonic beam approach has shown limitations since. Suitable chiral molecules for a parity violation measurement are solids with little to no vapour pressure~\cite{Stoeffler2011,Saleh2013}. They are thus poorly suited to a continuous supersonic beam setup which requires significant vapour pressure. Buffer-gas-cooled molecular beams formed using laser ablation of solid-state molecules in a cryogenic cell exhibit some of the highest beam fluxes to date, and are thus now considered~\cite{Tokunaga2017}. They also exhibit lower velocities, allowing for yet longer interaction times.}. For such experiments transit-time broadening is in fact a limiting factor and the use of trapped molecules would dramatically increase the interrogation time. Trapping and cooling molecules has been and is currently an enormous growth area at the interface between physics and chemistry. However, despite many key advances, there is currently no clear path to trapping large numbers of neutral molecules for long periods. Furthermore, the range of species adapted for trapping and cooling is limited\endnote{With the exception of a few dimers formed by the photo-association or magneto-association of pre-cooled atoms, the ultra-cold world was until recently confined to atomic systems. The extension of laser cooling to neutral molecules has recently begun, with the demonstration of magneto-optical traps of diatomic molecules~\cite{Barry2014,Truppe2017,Anderegg2017} and Sysiphus (optoelectrical~\cite{Prehn2016} or laser~\cite{Kozyryev2017}) cooling of polyatomic species, producing molecular samples at temperature down to 50~$\mu$K~\cite{Truppe2017}. In the last two decades several methods to control beams of gas-phase molecules (in particular decelerators of various kinds~\cite{Carr2009}) have been developed, leading to the first molecular fountain, producing ammonia molecules in free-fall for up to 266~ms~\cite{Cheng2016}, or to the trapping of samples of CO molecules on a microchip, at temperatures as low as 5~mK~\cite{Marx2013}. We finally note that increasingly complex molecules have been cooled to 1~K using buffer-gas cooling, exploiting collisions with cryogenically cooled noble gas atoms (see~\cite{Tokunaga2017} and references therein). We however stress again that the range of neutral species amenable to trapping and cooling is limited.}.

On the other hand, any ion, atomic or molecular, can be trapped indefinitely albeit in limited number. Trapped and laser cooled atomic ions have had an enormous impact in a number of fields from the development of future optical frequency standards (frequency metrology) to applications in quantum information processing (QIP)\endnote{Small ion ensembles can nowadays be controlled to form a truly programmable 5-qubit quantum computer~\cite{Debnath2016}.}. In 2009, the highest resolution spectroscopy ever performed had been achieved using ideas born at the interface between these closely related fields, leading to the emerging field of quantum metrology. Truly remarkable relative frequency resolution at the level of $5\times10^{-17}$ had been achieved by comparing ‘clock’ transition frequencies from two experiments, one in Hg$^{+}$ and the other in Al$^{+}$~\cite{Rosenband2008}\endnote{We note that, recently, a Yb$^+$ single ion clock achieved an unprecedented relative frequency uncertainty of $3\times10^{-18}$~\cite{Huntemann2016}.}. The Al$^{+}$ work was particularly exciting since it employed a radically new approach to ultra-high resolution spectroscopy – so called Quantum Logic Spectroscopy (QLS). This unprecedented accuracy had also led to new physics, setting stringent limits on the postulated time variation of the fine structure constant\endnotemark[\getrefnumber{mudot}].

Our proposed project aimed to use this new generation of ground-breaking spectroscopic techniques to probe trapped molecular ions in order to advance our understanding of a variety of fundamental processes in nature. In particular, we aimed to adapt the technique of Quantum Logic Spectroscopy to molecular ions\endnote{Since this proposal has been written, Quantum Logic Spectroscopy has been extended to ionic molecules, demonstrating the ability to non destructively measure a given molecular state~\cite{Wolf2016} and even manipulate it~\cite{Chou2017}.\label{qls}}. Our ultimate goals were new approaches to the observation of parity violation in chiral molecules and tests of time variations of the fundamental constants. Molecular ions also provide a route for measurements of the electron EDM and the techniques we aimed to develop are highly relevant to this approach~\cite{Meyer2008,Meyer2006}\endnotemark[\getrefnumber{edm}].

\section{Background}

\subsection{Ion trapping}
A linear radiofrequency (rf) ion trap consists of a set of four parallel rods with their centres at the corners of a square. An rf drive voltage is applied to one opposing pair of rods, while the other two rods are grounded. The applied voltage sets up a time averaged two-dimensional (radial) pseudopotential well in which ions may be trapped. Ions are prevented from escaping from the ends of the structure with a pair of `endcap' electrodes to which a dc potential is applied. The ions move harmonically in the trap with distinct motional frequencies along the various symmetry directions. The ions also experience a small, faster, motion at the trap drive frequency (the micro-motion). Ions trapped along the axis of the trap feel no micro-motion since the oscillating electric field is zero here. For a cloud of ions those ions furthest away from the centre line of the trap experience the greatest micro-motion~\cite{Blatt2008}.

\subsection{Laser cooling}
The application of laser cooling to trapped ions is particularly straightforward since the ions are already trapped\endnote{For a review of the various cooling techniques that can be applied to trapped ions we refer the reader to~\cite{Itano1995}. See also the recent demonstration of an electromagnetically-induced-transparency cooling method to cool to the ground state the radial motional modes of an 18-ion string~\cite{Lechner2016}.}. A single laser beam direction suffices, provided it has components along the major axes of the trap. Ion traps are much steeper (and deeper) than atom traps (typically $\omega_{\mathrm{trap}}\sim2\pi\times1$~MHz) so that the quantum states of the motion of the ion in the trap can be resolved if a narrow optical transition is employed. For very tight confinement, miniature traps can be employed. These traps allow operation in the Lamb-Dicke limit where the extent of the ground state motional wavefunction is less than $\lambda/2\pi$, where $\lambda$ is the wavelength of the excitation laser. In this limit, the spectrum of the ion is a carrier with a single pair of sidebands at the trap motional frequency. This makes optical sideband cooling the sub-Doppler cooling method of choice for trapped ions.

\subsection{State manipulation}
The internal state of an ion can be controlled by driving Rabi oscillations on a narrow `qubit' or `clock' transition between the ground state and a metastable state. Typically, the metastable state would be another hyperfine level of the ground state and the transition would be driven using two phase-coherent lasers in a Raman scheme. Laser pulses of known duration and intensity can be used to transfer population between states or to generate well controlled superpositions. The final state of an ion can be inferred using the `electron shelving' approach. One level of the `qubit' transition is connected to a short lived excited state (`read-out' level) such that fluorescence can be observed if the ion is in this level. The other level of the `qubit' is not connected to the short lived state so the ion appears `dark' under irradiation on the `read-out' transition. In this way, a superposition state is projected into one of the eigenstates. The amplitudes must be measured by repeated realisations of the projective measurement\endnote{The ion internal degrees of freedom manipulations described here have been introduced in the context of quantum information processing (QIP), see for example~\cite{Wineland2003}. QIP is still a very active field and nowadays, the trend is to perform those operations - except for detection - using direct microwave addressing of the qubit transition~\cite{Ospelkaus2011,Timoney2011,Lekitsch2017}, which is less prone to scattered photons. Very recently the same microwave-based  techniques were successfully applied to trapped molecular ions~\cite{Chou2017}.}.

\subsection{Sympathetic cooling}
In this process, two species of ions are loaded into a trap at the same time. One of the ions is chosen on the basis that a simple laser cooling scheme exists for it. The other ion can be anything from another isotope of the cooled species to an entirely different atomic ion or even a molecular ion. Cooling of these other ions is then mediated by long range Coulomb interactions between them and the laser cooled ions. It has been demonstrated that sympathetic cooling of molecular ions by laser cooled atomic ions efficiently lowers the translational energy of the molecular ions but in general leaves them vibrationally and rotationally hot. This is because the Coulomb interaction operates over such a long range and the partners do not approach each other closely enough for their internal degrees of freedom to be affected. A variety of approaches to the rotational and vibrational cooling of trapped molecular ions are currently being developed~\cite{Staanum2010,Schneider2010}\endnote{Internal state cooling via optical pumping using schemes related to those in~\cite{Staanum2010,Schneider2010} (in particular using broadband and incoherent optical sources) has been demonstrated for a variety of bialkali dimers as well as for a number of diatomic molecular ions (for a very good recent review, see~\cite{Hamamda2015}). The first implementation for a polyatomic molecule, methyl fluoride (CH$_3$F), has been reported recently~\cite{Glockner2015}. We also note that non optical internal state cooling has recently been demonstrated for BaCl$^+$~\cite{Rellergert2013} and MgH$^+$~\cite{Hansen2014} molecular ions, through sympathetic cooling with respectively ultra-cold calcium atoms (see also note\endnotemark[\getrefnumber{cooling}] below) or helium buffer-gas. \label{rovib cooling}}. Sympathetic cooling of a single molecular ion by a single laser cooled atomic ion has been demonstrated~\cite{Drewsen2004}. Two dissimilar atomic ions held together in a trap have been cooled to the motional ground state by applying laser cooling to one of the ions and depending on sympathetic cooling of the other ion~\cite{Barrett2003}\endnote{Sympathetic cooling relying on the Coulomb interaction between dissimilar ions is very efficient. For instance small ion crystals including molecular ions have been cooled to the motional ground state~\cite{Rugango2015}. Using an astute selective photo-ionization scheme combined with sympathetic cooling it has been possible to produce samples of rotationally and vibrationally state-selected, translationally cold molecular ions~\cite{Tong2010}. We also note that alternative strategies are promising: the reader will find a recent review of sympathetic cooling of molecular ions by ultra-cold neutral atoms in~\cite{Hudson2016}. \label{cooling}}.

\subsection{Quantum Logic Spectroscopy (QLS)}
Since QLS is at the heart of this proposal, we briefly outline it here. A full description can be found in reference~\cite{Schmidt2005}. Two atomic ions of different species are loaded into a linear radiofrequency ion trap. One ion is chosen for its ease of use in laser cooling, internal-state manipulation (creating superpositions of ground and excited electronic states) and measurement (measuring the final state by a quantum projection method with excellent fidelity). This ion, which we will call the `logic' ion, is used to sympathetically cool the other ion, so that the ion pair finds itself in the motional ground state of the trap potential. This removes many constraints from the choice of the other ion and it can be chosen almost exclusively on the basis of having a good candidate `clock' transition (a transition which has small systematic stray-field-dependent shifts). Determining the line shape of the `spectroscopy' ion proceeds in multi-step process. First, the spectroscopy ion is subjected to a laser pulse that creates some unknown superposition of the ground and excited state. The amplitudes of the superposition depend on the detuning of the exciting laser (which will eventually be scanned to provide the line shape). By the application of a second pulse with a specific power, duration and tuning to the spectroscopy ion, the initial superposition created in that ion is mapped onto the motional state of the ion pair. A final tailored pulse on the logic ion then maps the superposition onto the logic ion whence it can be measured using the `electron-shelving' technique described above.

One of the remarkable things about QLS is the freedom of choice conferred on the species of spectroscopic interest. In particular, the technique could be applied to molecular ions\endnotemark[\getrefnumber{qls}]. This opens up an enormous range of possibilities for measurements of features for which molecules are the system of choice. Ultra-high resolution molecular spectroscopy is currently used in a number of fundamental experiments, for instance searches for the electric dipole moment of the electron~\cite{Hudson2002}, the determination of the electron-to-proton mass ratio~\cite{Koelemeij2007,Korobov2009}, and the search for evidence of parity violation in chiral molecules~\cite{Daussy1999,Ziskind2002}\endnote{Precise spectroscopic measurements with molecules have indeed increasingly been used in the last decade to test fundamental symmetries (parity~\cite{Tokunaga2013,Asselin2017}, or parity and time-reversal, a signature of which would be the existence of a non-zero electron EDM, see note\endnotemark[\getrefnumber{edm}]) and postulates of quantum mechanics~\cite{CancioPastor2015}, to measure either absolute values of fundamental constants (such as the Boltzmann constant $k_{\mathrm{B}}$~\cite{Moretti2013,Mejri2015}, or the proton-to-electron mass ratio $\mu$, with recently the first determination of $\mu$ from a molecular system~\cite{Biesheuvel2016}), or their variation in time (fine structure constant $\alpha$~\cite{Hudson2006}, proton-to-electron mass ratio $\mu$, see note\endnotemark[\getrefnumber{mudot}])}.

\section{Work Plan}
We have opted to base our experimental proposal upon trapped Sr$^+$ ions. These ions will be used for sympathetic laser cooling of molecular ions and ultimately as the logic ion in QLS. This choice is based on a number of factors, not least of which is that it is a moderately heavy ion and is therefore well matched in terms of mass with the small to medium sized molecules we intend to study. A further advantage is that the SrH$^+$ molecule, which should be readily producible in the trap, may itself prove to be an excellent candidate for tests of the time-variation of the electron-to-proton mass ratio. Our initial experiments will be done with mixed clouds of Sr$^+$ and SrH$^+$ in a relatively large ion trap. We will initially create Coulomb crystals of Sr$^+$ ions, and demonstrate control over ion strings and even single ions. We will perform sympathetic cooling of Sr$^+$ isotopes and, by admitting H$_2$ gas to the chamber, will look for the production of sympathetically cooled SrH$^+$.

\begin{figure}
\centering
\includegraphics[width=\textwidth]{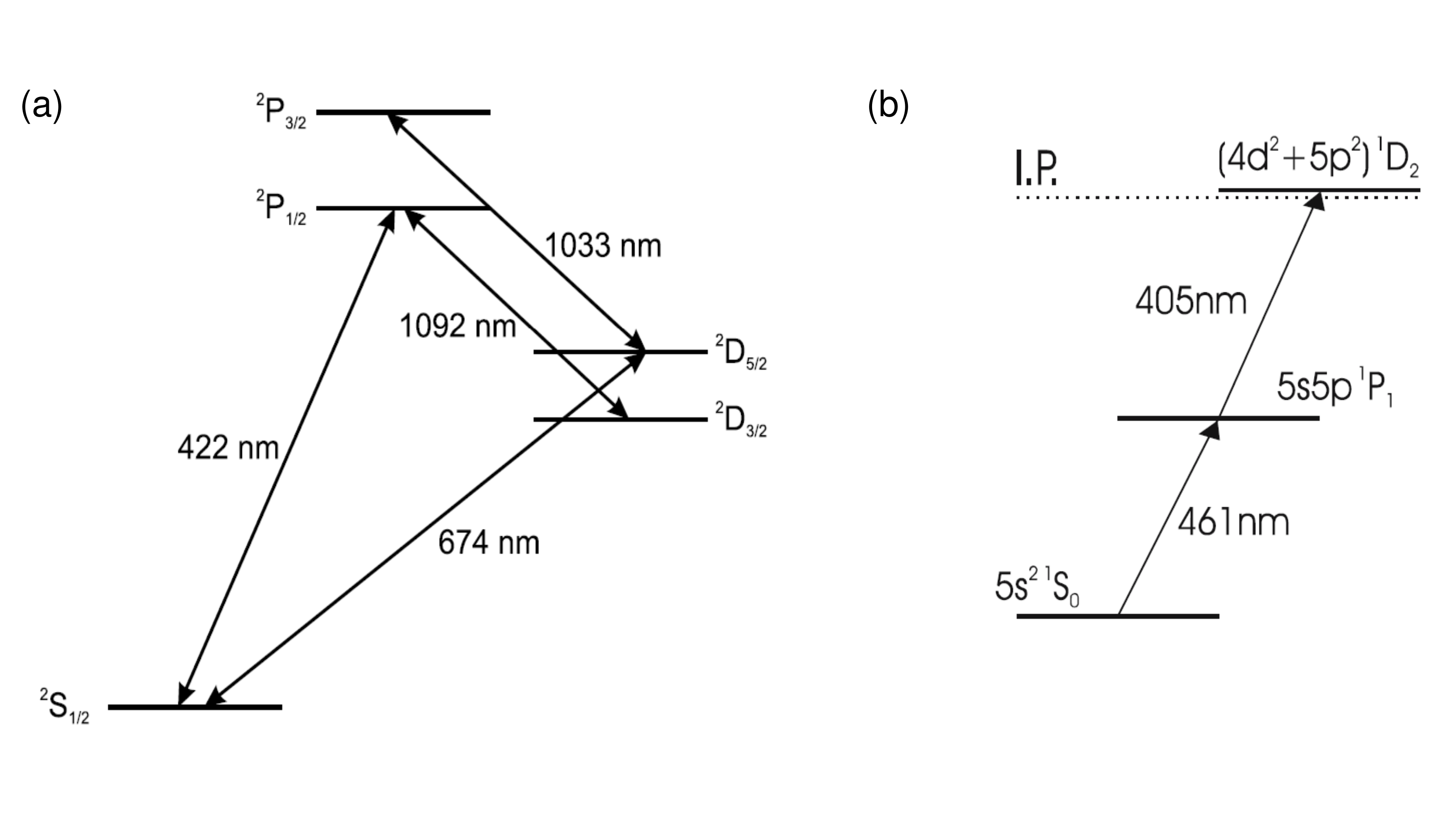}
\caption{(a) Relevant energy levels and wavelengths for Sr$^+$ (from ref.~\cite{Vant2006}). Laser cooling is via the 422~nm transition. Light at 1092~nm is needed to repump population that falls into the $^2$D$_{3/2}$ state. Light at 674~nm on the forbidden S-D transition is used for sideband cooling and quantum logic spectroscopy. Light at 1033~nm is used to speed up return from the $^2$D$_{5/2}$ state during sideband cooling. (b) Relevant transition wavelengths for photo-ionisation of neutral Sr (from ref.~\cite{Vant2006}). I.P.: ionisation potential.} \label{figure1}
\end{figure}

The laser systems required for cooling, repumping, and photoionisation is displayed in figure~\ref{figure1}. The 422~nm cooling laser will be a frequency doubled diode laser, whereas the 1092~nm laser will be a fibre laser. Light at 461~nm will be produced by an infrared diode laser frequency doubled in a single pass through a periodically poled second harmonic generation crystal. The final photoionisation step at 405~nm will be provided by a violet laser diode. The first generation ion trap will be a conventional linear radiofrequency trap designed to hold reasonably large ion crystals. A second generation trap will be required later to enable operation with single ions or small ion crystals in the Lamb-Dicke limit.

The various lasers need to be switched on and off rapidly for the various phases of the experiment. This will be done using a system of computer controlled shutters. A well designed, high efficiency, high resolution imaging system is also required for many of the experiments we envisage. The first objective will be to load and Doppler cool moderate sized clouds of Sr$^+$ ions (100's of ions). After this we will reduce the power driving the atomic beam oven until we reach the single ion regime. With a single ion in the trap, we will use a photon-rf correlation technique to reduce the micro-motion for ions held on the axis of the trap. Increasing the oven power will allow us to create small strings of ions and this will allow us to investigate sympathetic cooling, first of other Sr$^+$ isotopes and then of molecular ions. For the purposes of studying sympathetic cooling, a variety of molecular ions could be employed. However, we have an interest in working with SrH$^+$. Evidence from other work with Mg$^+$ and Ca$^+$ ions indicates that the production of SrH$^+$ is likely to happen naturally in the trap, at a slow rate~\cite{Mølhave2000,CaH+}. This proceeds via interactions with H$_2$ molecules present in the background gas. The rate of production of MgH$^+$ and CaH$^+$ has been enhanced by the deliberate admission of a small amount of H$_2$ gas into the vacuum chamber\endnote{Over the last few years, several techniques for the production of molecular ions have emerged and are reviewed in~\cite{Hudson2016}. The technique pioneered by the Drewsen group consists in first trapping an atomic ion of interest, usually produced by ionization of a neutral atomic gas, and then leaking in neutral molecular gas to react with the atomic ions and produce the desired molecular ions~\cite{Mølhave2000}. Recent examples are implementations by the Brown and Chapman groups to produce CaH$^+$~\cite{Khanyile2015} and Ba containing molecular ions~\cite{DePalatis2013} respectively. Another method for producing molecular ions is via laser ablation of a solid target~\cite{Hudson2016}. In a third method reviewed in~\cite{Willitsch2012}, a low-density neutral gas containing a parent molecule is leaked into the vacuum chamber and an electron beam or lasers are used to ionize the neutral molecules producing the desired molecular ion inside the trap volume.}. We will attempt to follow this procedure with Sr$^+$ ions. To confirm the presence of SrH$^+$ ions, we will perform \emph{in-situ} mass spectrometry by driving the trap with an additional small oscillating voltage resonant with a trap motional frequency for the molecular ion. The frequency of this drive is swept and when it hits a trap motional resonance, the ion crystal absorbs energy from the drive and changes shape. This leads to a change in the level of fluorescence.

For the experiments described thus far, Doppler laser cooling is sufficient. The next phase of the project will involve gaining control over the motional state of the trapped ions at the quantum level. This will require the development of more laser systems and a different ion trap. As illustrated in figure~\ref{figure1}, a highly stable 674~nm diode laser system is required to drive the narrow S-D logic ion transition, along with another infrared laser at 1033~nm used to speed up the sideband cooling process and to clear out the $^2$D$_{5/2}$ level.

In principle it would be possible to develop a small trap capable of operating in the Lamb-Dicke limit from the outset, but we feel it would be more prudent to approach this in two steps, given the somewhat greater technical challenges associated with operating small traps. The higher drive frequency of miniature traps requires a different means of generating the required rf voltage, employing a helical resonator. All of the required steps are now in reasonably wide use but there is still an overhead of effort required to implement them. In any case, traps of this variety are not well adapted to our initial experiments planned for larger ion clouds or crystals.

With the new trap in place, we would load a single Sr$^+$ ion, drive the S-D transition at 674~nm and observe Rabi Oscillations. With this level of control, we would be in a position to implement sideband cooling, first for a single ion and then for mixed crystals of either Sr$^+$ + SrH$^+$ or Sr$^+$ with some other molecular ion. Achieving this degree of control over few-ion crystals is a great technical challenge. Only a handful of groups worldwide have mastered all of the necessary techniques\endnote{Since then an ever increasing number of groups has mastered those techniques that are now part of the `toolbox' to manipulate trapped ions. Dissemination of this know-how is particularly important in the context of experiments on cold molecular ions. For instance the European Commission (FP7) supports since 2013 a Marie Curie initial training network called `Cold Molecular Ions at the Quantum limit' dedicated to this task (see http://itn-comiq.eu).}. With these developments complete, we will be in a position to perform quantum logic spectroscopy with a molecular ion as the spectroscopy ion. If possible we will use SrH$^+$ as our test ion, however little is known about this molecule apart from a calculation of its vibrational frequency in the electronic ground state~\cite{Schilling1987}\endnote{Recently, SrH$^+$ has attracted much interest in fields ranging from precision measurements to cold chemistry or astrophysics and several theoretical studies~\cite{Abe2010,Aymar2012,Habli2015,Belayouni2016} were reported for the calculation of the potential energy curves, permanent and transition dipole moments, static dipole polarizabilities of electronic states of SrH$^+$.}. Calculations have been performed for the lowest lying excited states of MgH$^+$~\cite{Jørgensen2005} and it is our intention to perform similar calculations for SrH$^+$. One significant difference is that the atomic D-states that exist in Sr$^+$ but not in Mg$^+$ should correlate with a further family of interatomic potentials for SrH$^+$ not present in MgH$^+$. It may then be possible to find a narrow (forbidden) $\Sigma$-$\Delta$ electronic ro-vibrational band and to use a line within this band to perform quantum logic spectroscopy in the first instance.

Failing this, another approach would be to use a direct ro-vibrational line for the quantum logic spectroscopy. The vibrational frequency of SrH$^+$ has been calculated as being 1346~cm$^{-1}$ leading to a wavelength of 7.43~$\mu$m for the $v=0$ to $v=1$ vibrational band~\cite{Schilling1987}. This transition would be amenable to a quantum cascade laser (QCL). The \emph{MMTF} group has a wealth of expertise in performing spectroscopy in this spectral region, mostly using CO$_2$ lasers at 10~$\mu$m and more recently QCLs\endnote{Since then, the \emph{MMTF} group has gained experience in frequency stabilizing QCLs at unprecedented levels and has used them to perform precise spectroscopic measurements~\cite{Tokunaga2017,Asselin2017,Sow2014,Argence2015}. The use of QCLs allows the study of any species showing absorption between 10 and 100~THz, paving the way for precise measurements on a considerably larger number of species.}.

It is important to note that, while for light ions like SrH$^+$ most of the population resides in the vibrational ground state at room temperature, the ions are distributed over a range of around 10 rotational states. In a single ion experiment, one could simply try the experiment many times until a molecule in the target rotational state is trapped. To improve the efficiency of the experiment, it would clearly be a great advantage to start with all the molecules in a well defined rotational level, preferably the ground state. There has been a great deal of activity in the general area of ground-ro-vibrational state cooling of molecules recently, for both ions and neutral systems. In the last year, two groups have reported significant progress in active laser cooling of trapped molecular ions to the ro-vibrational ground state~\cite{Staanum2010,Schneider2010}\endnotemark[\getrefnumber{rovib cooling}].

Other approaches to ground state molecular cooling of molecular ions include using ultra-cold atoms for sympathetic cooling\endnotemark[\getrefnumber{rovib cooling}]$^,$\endnotemark[\getrefnumber{cooling}], or simply building the ion trap in a cryostat. Cryogenic ion traps are not a new idea~\cite{Poitzsch1998} but recent developments in trap fabrication processes have meant that their use seems likely to become more widespread over the next few years. Indeed cryogenic ion traps are currently under development in a number of laboratories worldwide~\cite{Labaziewicz2008}\endnote{Recently, the ability to operate micro-fabricated planar silicon-based ion traps in a cryogenic environment was demonstrated~\cite{Niedermayr2014}, thus combining a scalable trap design with a low heating environment.}.

It is important to note that SrH$^+$ is not simply a `toy' molecule, useful only as a means of demonstrating quantum logic spectroscopy for a molecular ion. On the contrary, precision spectroscopy of this molecule would allow a stringent direct and model-free test of the time variation of the electron-to-proton mass ratio at the level of one part in $10^{15}$, which would be approximately one order of magnitude better than the current limit set by the \emph{MMTF} group using neutral SF$_6$~\cite{Shelkovnikov2008}\endnotemark[\getrefnumber{mudot}]$^,$\endnote{See for instance the work by M. Kajita \emph{et al.} who show that ultra-precise vibrational spectroscopy of SrH$^+$ ions is possible at the $10^{-16}$ level and can thus be used to probe the temporal stability of $\mu$ in the present epoch~\cite{Kajita2011,Kajita2014}, or could even be exploited in new generation molecular clocks~\cite{Kajita2016}.}.

Concurrently with the experimental developments described above, we will undertake a feasibility study for the application of these techniques to the measurement of parity non-conservation shifts in chiral molecules. The first task will be to identify a suitable chiral molecular ion. Preliminary discussions with quantum chemists  indicate that molecular ions may have some advantages over neutral chiral molecules for these studies in that the parity violation energy differences may be larger for ionic systems. One clear advantage of ions in this regard is that the sample of chiral molecules required is tiny since the method we will adopt is intrinsically a single-particle approach.

Performing quantum logic spectroscopy on a pair of chiral molecules could be done in a number of ways. The most obvious approach would be to perform experiments on different enantiomeric ions one at a time and to determine their central frequencies as accurately as possible. Since we are only interested in making relative measurements, certain systematic shifts may be less troublesome for this experiment than for frequency standards applications. Novel approaches based upon the use of entangled states~\cite{Roos2006,Chwalla2007} suggest a different protocol may be appropriate. This protocol would involve loading two molecular ions (one of each enantiomer) plus one logic ion into the trap at the same time. The techniques of quantum logic spectroscopy could then be used to first entangle the two molecular ions with each other. If the ions are prepared in an appropriate Bell state, a phase will accrue over time between the different parts of the Bell state wave function. This phase can then be read out by mapping the Bell state onto the collective motional state of the three-ion crystal. Finally, the motional state can be mapped onto the logic ion which can then be read out in the normal way. Oscillations in the final state of the logic ion at the difference frequency of the two molecules should then be apparent. This rather direct approach to measuring the energy gap between the molecules may have some advantages in terms of rejecting common mode noise and systematic effects.

Finally we note that the group of Eric Cornell is currently developing an experiment to measure the EDM of the electron using molecular ions~\cite{Stutz2004}\endnotemark[\getrefnumber{edm}].  The kind of spectroscopy we envisage developing may also have a significant impact on experiments of this kind.

We are convinced that a host of new spectroscopic techniques based upon ideas that have grown out of the work on quantum information processing is set to have a major impact on the general field of ultra-high precision measurements. We are therefore persuaded that embarking on an experimental programme in this general area is extremely timely.

\section*{Acknowledgements}

The authors thank H\'{e}l\`{e}ne Perrin, Anne Amy-Klein, Olivier Dulieu, Trond Saue, Jeanne Crassous and the French ion trapping community - in particular the members of the \emph{Quantum Information and Technologies} group from the Laboratoire Matériaux et Phénomènes Quantiques, Universit\'{e} Paris-Diderot; the members of the \emph{Trapped Ions} group from the Laboratoire Kastler Brossel, Universit\'{e} Pierre et Marie Curie; and the members of \emph{Confinement d'Ions et Manipulation Laser} group from the Physique des Interactions Ioniques et Moléculaires laboratory, Aix-Marseille Universit\'{e} - for fruitful discussions. We would also like to thank the referees and editors for having very positivity considered and criticised this unconventional manuscript.

\section*{Funding}

This work was supported by the Universit\'{e} Paris 13 through the Invited Professor Programme; Agence Nationale de la Recherche under Grant ANR-15-CE30-0005-01; and R\'{e}gion \^{I}le-de-France through DIM Nano-K.

\theendnotes

\bibliographystyle{tfp}
\bibliography{danny}

\end{document}